\documentclass[12pt,draftcls,onecolumn]{IEEEtran}
\IEEEoverridecommandlockouts
\usepackage{cuted}
\usepackage{cite}
\usepackage{bm}
\usepackage{amsmath,amssymb,amsfonts}
\usepackage{algorithm}
\usepackage{algorithmic}
\usepackage{subfig}

\usepackage{graphicx}
\usepackage{makecell}
\usepackage{diagbox}

\usepackage[letterpaper, left=0.625in, right=0.625in, bottom=1.02in, top=0.70in]{geometry}

\usepackage{textcomp}
\usepackage{xcolor}

\newcommand{\be}{\begin{equation}}
\newcommand{\ee}{\end{equation}}

\def\BibTeX{{\rm B\kern-.05em{\sc i\kern-.025em b}\kern-.08em
    T\kern-.1667em\lower.7ex\hbox{E}\kern-.125emX}}

\begin{document}

\title{Spatial Channel State Information Prediction with Generative AI: Towards Holographic Communication and Digital Radio Twin}

\author{\IEEEauthorblockN{Lihao Zhang\IEEEauthorrefmark{1}, \emph{Student Member, IEEE}, Haijian Sun\IEEEauthorrefmark{1}, \emph{Member, IEEE}, Yong Zeng\IEEEauthorrefmark{2}, \emph{Senior Member, IEEE}, Rose Qingyang Hu\IEEEauthorrefmark{3}, \emph{Fellow, IEEE}} \\
\IEEEauthorblockA{
\IEEEauthorrefmark{1}School of Electrical and Computer Engineering, University of Georgia, Athens, GA, USA \\
\IEEEauthorrefmark{2}National Mobile Communications Research Laboratory, Southeast University, Nanjing, China. \\
\IEEEauthorrefmark{3}Department of Electrical and Computer Engineering, Utah State University, Logan, UT, USA \\
Emails: \{lihao.zhang, hsun\}@uga.edu}, yong\_zeng@seu.edu, rose.hu@usu.edu}

\maketitle

\begin{abstract}
As 5G technology becomes increasingly established, the anticipation for 6G is growing, which promises to deliver faster and more reliable wireless connections via cutting-edge radio technologies. However, efficient management method of the large-scale antenna arrays deployed by those radio technologies is crucial. Traditional management methods are mainly reactive, usually based on feedback from users to adapt to the dynamic wireless channel.  However, a more promising approach lies in the prediction of spatial channel state information (spatial-CSI), which is an all-inclusive channel characterization and consists of all the feasible line-of-sight (LoS) and non-line-of-sight (NLoS) paths between the transmitter (Tx) and receiver (Rx), with the three-dimension (3D) trajectory, attenuation, phase shift, delay, and polarization of each path. Advances in hardware and neural networks make it possible to predict such spatial-CSI using precise environmental information, and further look into the possibility of holographic communication, which implies complete control over every aspect of the radio waves emitted. Based on the integration of holographic communication and digital twin, we proposed a new framework, digital radio twin, which takes advantages from both the digital world and deterministic control over radio waves, supporting a wide range of  high-level applications. As a preliminary attempt towards this visionary direction, in this paper, we explore the use of generative artificial intelligence (AI) to pinpoint the valid paths in a given environment, demonstrating promising results, and highlighting the potential of this approach in driving forward the evolution of 6G wireless communication technologies.

\end{abstract}


\section{Introduction}
\par Wireless communication system exchanges information via electromagnetic (EM) waves  between the physically separated Tx and Rx. These EM waves, as traversing a three-dimensional (3D) environment from the Tx, interact with objects, undergo reflection, refraction and diffraction, and may arrive at the Rx through different paths. These paths lead to variations in time-of-flight, attenuation, polarization, and phase at each of them and result in distorted signals at the Rx, collectively known as the multi-path effect. To mitigate this, techniques such as equalization are commonly employed, which treat the wireless channel as time variant or invariant systems within its coherence time, periodically estimate the channel state information (CSI) at the Rx, and use various techniques to restore the signal. However, this method faces significant limitations in terms of high overhead and latency, and demands for sophisticated receiver structures in the more intelligent and dynamic next-generation wireless systems. 


\par There is a notable trend towards actively leveraging the \emph{spatial} capabilities of wireless channels. For example, massive multiple-input multiple-output (mMIMO) \cite{mMIMO_lu2014overview}, characterized by a substantial number of antennas, facilitates beamforming schemes at the Tx to more effectively exploit the existing spatial wireless resources. As another example,  reconfigurable intelligent surfaces (RISs) \cite{RIS_liu2021reconfigurable} that comprise vast arrays of dynamically reconfigurable EM elements, can modify the characteristics of incoming EM wave, such as amplitude, phase, polarization, and reflection direction. This capability enables the generation of more spatial wireless resources by creating additional ``virtual line-of-sight'' (LoS) paths. 

\par Beyond the specific radio technologies, a visionary concept is holographic communication \cite{holo_huang2020holographic}.  Similar to holographic imaging, holographic communication aims to advance towards comprehensive control of EM waves, anticipated to be realized through the cooperation of mMIMO and RIS, enabling deterministic spatial beam management from the Tx to the Rx. Collaborating with the well-established temporal and spectral control of wireless channel, this ambitious concept promises to provide ultra-fast, stable and predictable wireless links even in intricate environments, enhancing the support for high-level applications, such as digital twin and connected and autonomous vehicles (CAVs).

\par Previous studies on these innovative concepts have primarily relied on receiver feedback to formulate the precoding matrix or beam pattern. Yet, in terms of the feedback, the receivers, usually end-users, possess limited capabilities to obtain the accurate channels and provide a comprehensive feedback, especially for the large arrays. Regarding formulating the management schemes, it requires considerable effort in modeling the problem and developing optimization algorithms, and its effectiveness is often constrained by the inconsistent wireless channel.  For example, in small-scale MIMO systems, the Tx and Rx use predefined codebook for sub-optimal precoding. In mMIMO systems, the large number of antennas leads to widespread use of beam scanning to identify the most effective beamforming pattern. For RISs, which work as remote antennas to the base station (BS), it is also more efficient to manage them with beamforming, considering the vast number of their EM elements. Holographic communication further takes into consideration the propagation from RISs to the users, where the overhead is expected to be extremely high to manage the remotely located RISs.

\par Essentially, performance gain of these advanced radio technologies can be enhanced by the accurate knowledge of \emph{spatial-CSI}, which consists of all the significant LoS and non-LoS (NLoS) paths between the Tx and Rx, with the 3D trajectory, attenuation, phase shift, delay, and polarization of each path. Accordingly, with real-time spatial-CSI  prediction,  we can explicitly manipulate the beams in the 3D space with beamforming technology to align with these paths \cite{zeng2023tutorial}. Within the context of digital radio twin, the tasks of spatial-CSI prediction and scenario sensing (sensing the environment and the moving objects) can be collaboratively integrated. For instance, the metallic surface of a specific vehicle can alter the spatial-CSI to a certain extent, and the discrepancies between prediction and measurement can be leveraged for sensing the vehicle. Then, with the sensing data, we can update the digital radio twin system for subsequent spatial-CSI predictions.

\par Nevertheless, to date, there still lacks a practical method for real-time prediction of spatial-CSI.  Currently, one viable approach to acquire spatial-CSI is through radio simulation, which faces two challenges: the cost of gathering and processing precise environment information, and the computational complexity involved in the simulations. On one hand, thanks to the ongoing validity of Moore's law and the significant contributions of researchers and engineers in hardware development, there has been an exponential increase in computational capability at similar prices and power consumption. This trend is also observed in storage devices and sensors, facilitating the collection, processing, and storage of more comprehensive and accurate environment data at an acceptable cost. On the other hand, the rapid advancements in the field of AI have led to the development of numerous powerful neural networks. These advancements have prompted the acceleration of radio propagation simulations using neural networks. Several innovative attempts have been made in this direction, including the construction of 3D environments with electromagnetic properties derived from images \cite{mmsv_kamari2023mmsv}, calibration of electromagnetic properties using differentiable radio simulators \cite{sionna_ruah2023calibrating}, and neural 3D ray tracing \cite{winert_orekondy2022winert}.

\par Generative AI, a rapidly growing and highly-regarded branch of AI, is particularly notable for its capacity to process and creatively output multimodal content \cite{gene_AI_harshvardhan2020comprehensive}. This paper investigates how generative AI can be applied for accurate spatial-CSI prediction. In the phase of constructing environment, generative AI enables integrating multi-modal data sources (images, point clouds, text labels, and numerical data) to reconstruct the 3D environment with visual and electromagnetic material properties and detail it by prior knowledge. For example, we can obtain a car's make and model by semantic identification from images, and then load the corresponding stored 3D model and materials as a reference for further processes. Radio simulation in the 3D environment begins with a ray launching process which emits numerous rays ( $\geq 100,000$), then traces each of them until they either reach the number of bounces limits or their intended destinations. This approach can be seen as a method of exhaustive environmental sampling, especially considering that the number of valid paths typically does not exceed 200 in enclosed indoor environment and 50 in open city environment. Traditional neural networks face challenges in directly learning the complex main paths, such as number of paths, number of bounces of each path, and match up of paths. Generative AI, however, shows greater capability in processing multi-modal environmental information input and output complex content. This paper presents an initial exploration into neural radio tracing using an image-to-image generative network.

\par The contributions of this paper are summarized as follows:
\begin{itemize}
\item We examine the emerging 6G radio technologies, focusing on their exploration of spatial wireless resources and the advantages they can derive from spatial-CSI.
\item We introduce the concept of digital radio twin, distinguished by its emphasis on spatial-CSI prediction and the facilitation of holographic communication.
\item We investigate the application of neural radio tracing using generative AI, aiming to accelerate radio simulation for real-time spatial-CSI prediction.
\item We identify and discuss several key challenges that are critical to the future development of holographic communication and the digital radio twin systems.
\end{itemize}

\section{Background and Motivation}

\subsection{Wireless Channel Prediction}
\par As mentioned earlier, the predominant method for acquiring channel information is pilot-signal-based channel estimation, which can serve as a prediction for only the next few transmissions with a lack of generalization capability. Consequently, many statistical channel models have been proposed based on large amount of field measurement, which can effectively predict signal strength levels based on Tx-Rx distance and empirical parameters for different environments. However, statistical models fall short in providing the certainty and spatial information about its channels, which are crucial for holographic communication.

\par In contrast, deterministic channel models are grounded in detailed environmental information and EM theory to describe the propagation of EM waves. Among these models, computational electromagnetic methods (CEMs) are the most solid ones, solving Maxwell's equations with detailed boundary conditions through computational methods like the finite element method and finite-difference time-domain \cite{CEM_bondeson2012computational}. These methods offer high accuracy and are particularly effective in small-scale environments, but become increasingly challenging and less practical in large-scale environments due to the immense computational demands.

\par Considering that the average size of objects and environments in large-scale environments is substantially larger than cellular EM wavelength, most space can be regarded as the far-field relative to the Tx, and it is feasible to employ the ray concept from geometric optics as an approximation to the propagation of EM wave \cite{GO_born2013principles}. This approximation has led to the widespread adoption of ray tracing methods for radio wave simulation, offering a balance between computational efficiency and accuracy in complex, large-scale environments. 

\begin{figure*}[ht]
    \centering
    \includegraphics[width=1.0\linewidth]{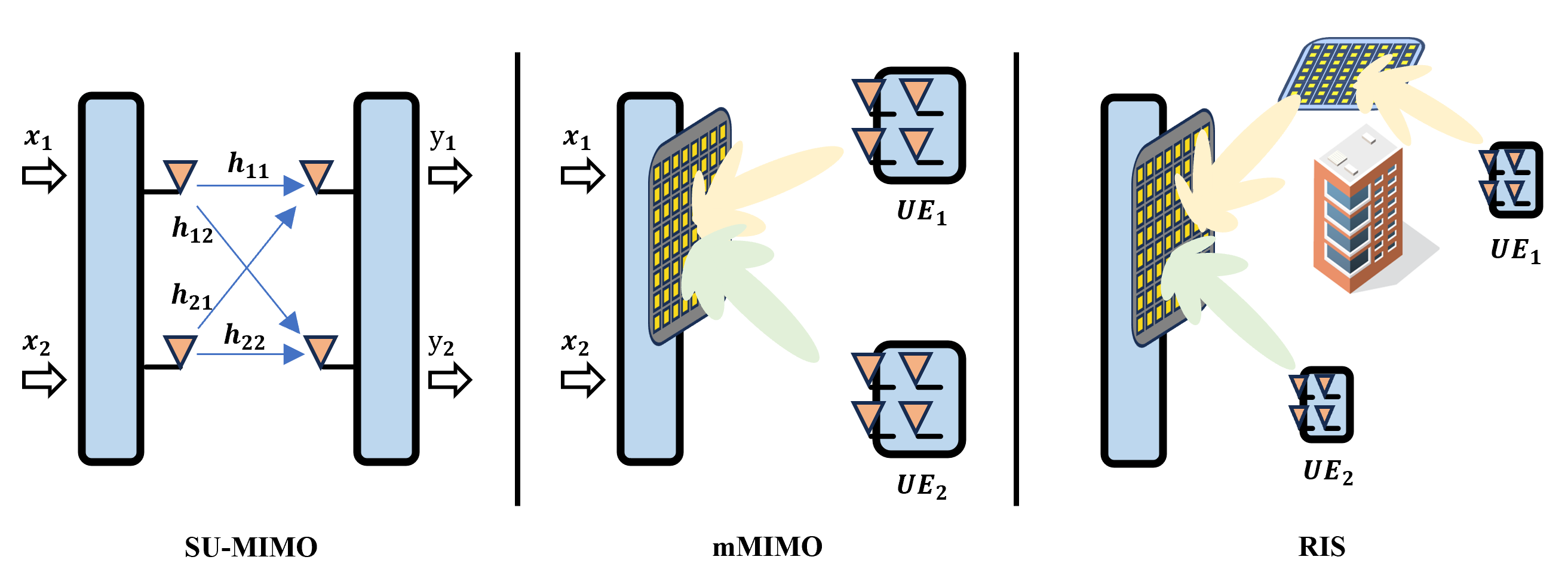}
    \caption{Radio technologies that exploit spatial multiplexing: SU-MIMO, mMIMO, and RIS. }
    \label{fig:6g_radio}
\end{figure*}

\subsection{Current Spatial-CSI Use Cases}

\subsubsection{SU-MIMO}

\par The advancement in MIMO systems represents a significant step in actively utilizing wireless channels. It is instructive to begin with a simplified $2\times2$ single-user MIMO (SU-MIMO) system, as depicted in Fig. \ref{fig:6g_radio}. In this setup, the complex channel coefficients between the Tx and Rx antennas form the channel matrix $H$. The Tx sends two different data streams to the receiver for higher throughput, known as spatial multiplexing scheme. Alternatively, in diversity scheme, the same data stream is transmitted through multiple antennas for enhanced reliability. 

\par To mitigate  mutual interference, a direct solution is precoding, which manipulates the data streams to make them orthogonal on the spatial streams. This process is implemented using the singular value decomposition of $H$, expressed as $H = U\Sigma V^H$, where $V$ is the precoding matrix. The non-zero diagonal elements in $\Sigma$ result in independent spatial streams. While the Rx can calculate and report channel back to the Tx, it poses a substantial overhead. A compromise involves maintaining a codebook on both sides, and the Rx only reports to the Tx which precoding matrix from the codebook should be utilized.

\subsubsection{mMIMO}
\par While small-scale MIMO system can implicitly utilize multi-path effects for spatial multiplexing gain, the number of effective spatial streams depends on the number of antennas, RF chain, and orthogonal channel paths. To address this, mMIMO systems was proposed to make the most of existing paths. As shown in Fig. \ref{fig:6g_radio}, mMIMO can form extremely narrow radio beams with enough antennas and RF chains, targeting only the valid LoS and NLoS paths to multiple users. By doing so, mMIMO can significantly boost system throughput by serving multiple users with identical spectral and temporal resources while providing numerous spatial streams per user. Another motivation of mMIMO system development is the exploitation of the millimeter-wave band in 5G and 6G networks, which lead to a reduction in antenna size and compact size of large-scale antenna arrays.

\subsubsection{RIS}
\par A natural approach to enhance the spatial wireless capacity is putting some ``mirrors" to construct more valid paths in current environment, and this is where RISs come into play. RIS is capable of not just fully reflecting, but also actively reshaping the beam with large number of EM elements, and can cooperate with beamforming as shown in Fig. \ref{fig:6g_radio}. However, the development of practical RIS devices still requires lots of effort in both hardware and beam management. In terms of hardware, numerous systems have already been actualized through substantial efforts \cite{RIS_1_zeng2023ris}. But for management, it also faces the similar challenge to mMIMO.

\subsubsection{Holographic Communication}
\par As holographic communication dedicating to establishing comprehensive control of the EM wave propagation from the Tx to Rx, it emphasized the management of mMIMO and RISs and efficient acquisition of channel information between the mMIMO BS, RISs, and users. Along the traditional workflow, the steps will be individually estimating the channels from the feedback, calculating the precoding matrix or beam patterns and transmitting these to the RISs in real-time either wirelessly or via wired connections. However, this workflow adds significant hardware costs and system complexity for the additional communications between BS and RISs, also substantial processing demands at the BS side considering the problem scale. As they are situated within the same local environment, using the BS with intelligent infrastructures to sense and understand the environment and predict the spatial-CSI would be more efficient.
 
\subsection{Spatial-CSI Prediction }
\par The explosive advancement in AI fields and the decreasing costs of computation and sensing hardware, are bringing us closer to holographic communication with real-time spatial-CSI prediction. Spatial-CSI, which is distinct from traditional CSI due to its detailed attributes of each path as mentioned before, enables the deterministic construction of spatial streams. This means that data streams can be precisely mapped to specific radio beams, which are then finely steered towards valid LoS and NLoS paths. Meanwhile, the receivers adjust their receiving beam patterns to align with these transmitted beams, facilitating interference-free reception. Moreover, as the spatial resolution of the system increases and more RISs are deployed to augment spatial wireless capacity, the benefits gained from holographic communication systems are expected to escalate rapidly. Therefore, the prediction of spatial-CSI, while a stiff and distant challenge, holds immense potential for the advancement of future wireless communication systems.

\section{Digital Radio Twin}

\par The process of sensing and understanding the local environment for spatial-CSI prediction in real-time highly coincide with the digital twin \cite{10148936}, which involves creating a digital replica of the physical world, capturing essential and detailed information about various entities and environments. This replica serves a crucial role in the sensing, analyzing, predicting, and planning of specific entities. Digital radio twin is a specialized form of digital twin, characterized by its description of spatial-CSI and the enabling of holographic communication, thereby elevating throughput, and ensuring consistent and predictable quality of service for each wireless link.
\begin{figure*}[ht]
    \centering
    \includegraphics[width=1.0\linewidth]{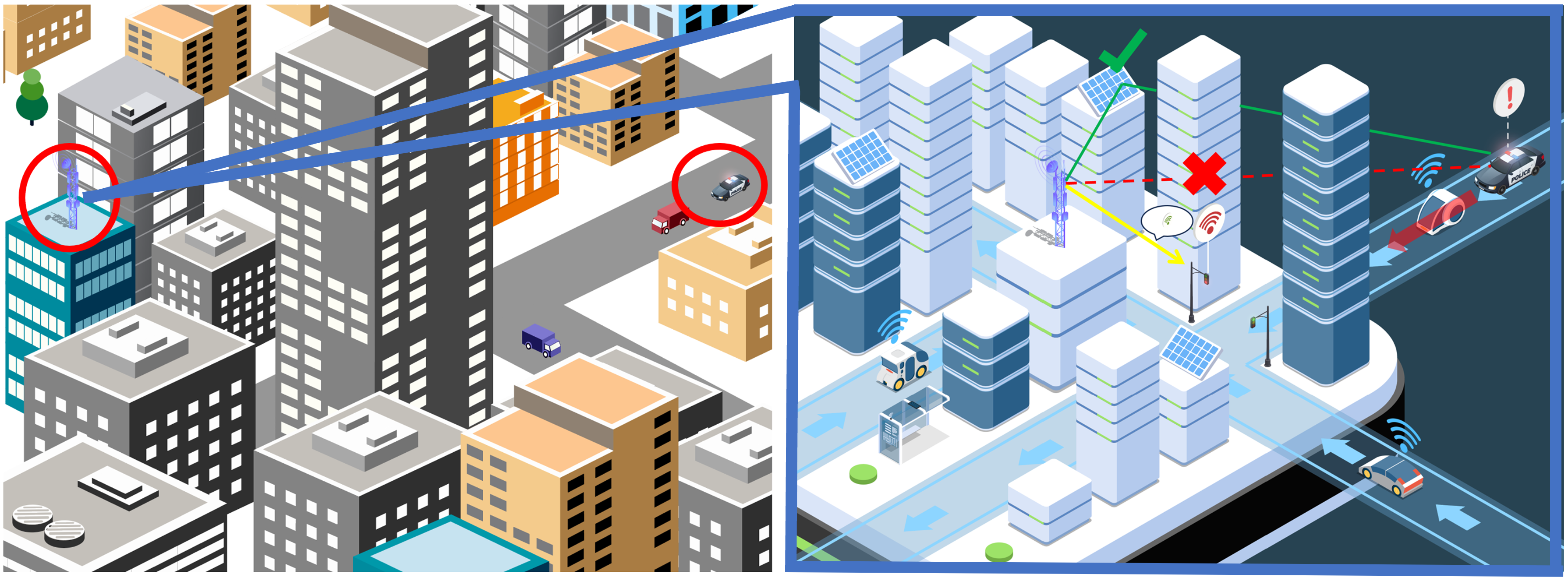}
    \caption{Digital radio twin operating in a local intelligent base station}
    \label{fig:digital_twin}
\end{figure*}
\subsection{Building Environment and Basic Radiomap}
In urban environments, where changes occur slowly and planned, it is practical to address these long-term elements separately. As depicted in Fig. \ref{fig:digital_twin}, most of computational tasks of digital twin are implemented within a local smart BS that governs a designated surrounding area. This involves constructing a digital environment, consisting of precise 3D models of the buildings, infrastructures and roads, incorporating visual textures, material segmentation, EM parameters, and textual labels (such as roads, sidewalks, and crossings), similar to the workflow proposed in \cite{mmsv_kamari2023mmsv}. Accompanying these long-term elements, middle-term elements, such as weather, temperature, and atmospheric scattering, should be updated regularly, though not in real-time. Upon completion of the basic digital environment, the BS can simulate and maintain a basic 3D radiomap which provides the spatial-CSI between the BS and all the positions in the 3D space. This map serves as an initial channel prediction for further processes.

\subsection{Alignment with Physical World}
\par The scheme of separately addressing short-term elements is based on two key considerations.  First, these elements, such as pedestrians and vehicles, are relatively small compared to the overall environment, and do not significantly alter the radio paths. Second, many of these short-term elements are of interest in higher-level applications, such as connected vehicle systems or pedestrian monitoring. At this point, the tasks of calibrating their impact on wireless channels and sensing these targets are organically integrated. For instance, a system may detect the entry of a connected vehicle into its range through variations in the wireless channel, subsequently sensing and integrating this vehicle into the digital twin, thereby updating channel predictions. Consequently, the prevailing challenge is the precise sensing and integration of these elements.

\par To solve this challenge, intelligent infrastructures, which are directly linked to the BS and equipped with various sensors, can gather a large amount of multi-modal data. Processing and fusion of these multi-modal data have been extensively explored, particularly within the field of CAVs. Beyond the data handling, from a systemic perspective, the sensing targets are classified into connected and unconnected entities. Connected entities, like the CAVs, can be seamlessly integrated into the digital radio twin system through several transmissions, and can continuously report their states, such as velocities, positions, directions and routes, to the digital radio twin system. For unconnected entities, such as bicyclists, the system needs to spend more effort to detect and sense them. This task can be done by collecting the data from sensors and processing this data with the digital radio twin. Besides, those connected entities, which usually possess various sensors, can provide additional data from completely different views.


\section{Neural Radio Tracing}
\par Based on the previous discussion, we can see the importance of spatial-CSI prediction for the next generation wireless system, especially its physical layer technologies. The current most feasible workflow for predicting spatial-CSI is radio simulation, and ray tracing is its time-consuming and computation-intensive step. Ray tracing involves emitting of a substantial number of rays, meticulously tracking their propagation, and ultimately identifying all the valid ray paths. Then, the channel characteristics are calculated and transformed into the baseband equivalent channel or the channel gain. While the use of advanced GPUs and CPUs has facilitated this process with compromised  accuracy, ray tracing remains excessively time-consuming for real-time applications in wireless communication. A critical consideration here is that the traditional shooting and bouncing ray tracing method is initially designed for image rendering, which is similar but also very different from radio simulation in some aspects.

\subsection{Image Rendering Versus Radio Simulation}

\par Image rendering primarily focuses on visible light, with spectrum ranging from 430 THz to 750 THz. During the propagation, these lights interact with objects and give rise to a large number of paths which will collectively influence the imaging fidelity. Rendering high-quality image demands exceptional pixel quantity, and multiple rays per pixel (for instance, rendering at $1920 \times 1080$ pixels with 10 rays per pixel requires 20,736,000 rays). Fortunately, human vision is less demanding in terms of spectral and temporal resolution, which only utilizes 24 bits (r, g, b with 8 bit) via a 300 THz bandwidth on each pixel, and 24 frames per second. 

\par In contrast, radio simulation presents a considerable different scenario. Although the radio wavebands are commonly discussed, they actually utilize discrete frequencies evenly distributed within the designated waveband. Contrary to the continuous and mixed spectrum of certain specific color, a significant effort in wireless systems is dedicated to mitigating the interference between these discrete frequencies. The propagation characteristics of radio waves also diverge substantially from light, as material EM properties vary greatly between sub-6G and 500 THz and  only reflection and diffraction play pivotal roles. Finally, at the Rx, radio simulation requires extremely high resolution on frequency and time domain but very little in space compared to image rendering.

\par  From this comparison, it is apparent that there is a need to optimize the ray tracing methods for radio simulation, termed as ``radio tracing''. The challenges in radio tracing primarily reside in efficiently identifying the geometrical paths, and precisely modeling and calibrating the EM wave propagation process. Unlike image rendering, where the vast majority of emitted rays contribute to the final image quality, in radio simulation, the number of valid paths rarely exceeds 200, suggesting a more targeted approach in identifying these paths. Once the accurate geometrical paths are determined, the next challenge is calculating the precise channel characteristics. Contrary to image rendering, which focuses on the 24-bit color information of each pixel, radio simulation requires accuracy in attenuation, phase shift, delay, and polarization, necessitating considerable effort in more refined modeling method, field measurement, and calibration methods.

\begin{figure*}[ht]
    \centering
    \includegraphics[width=1.0\linewidth]{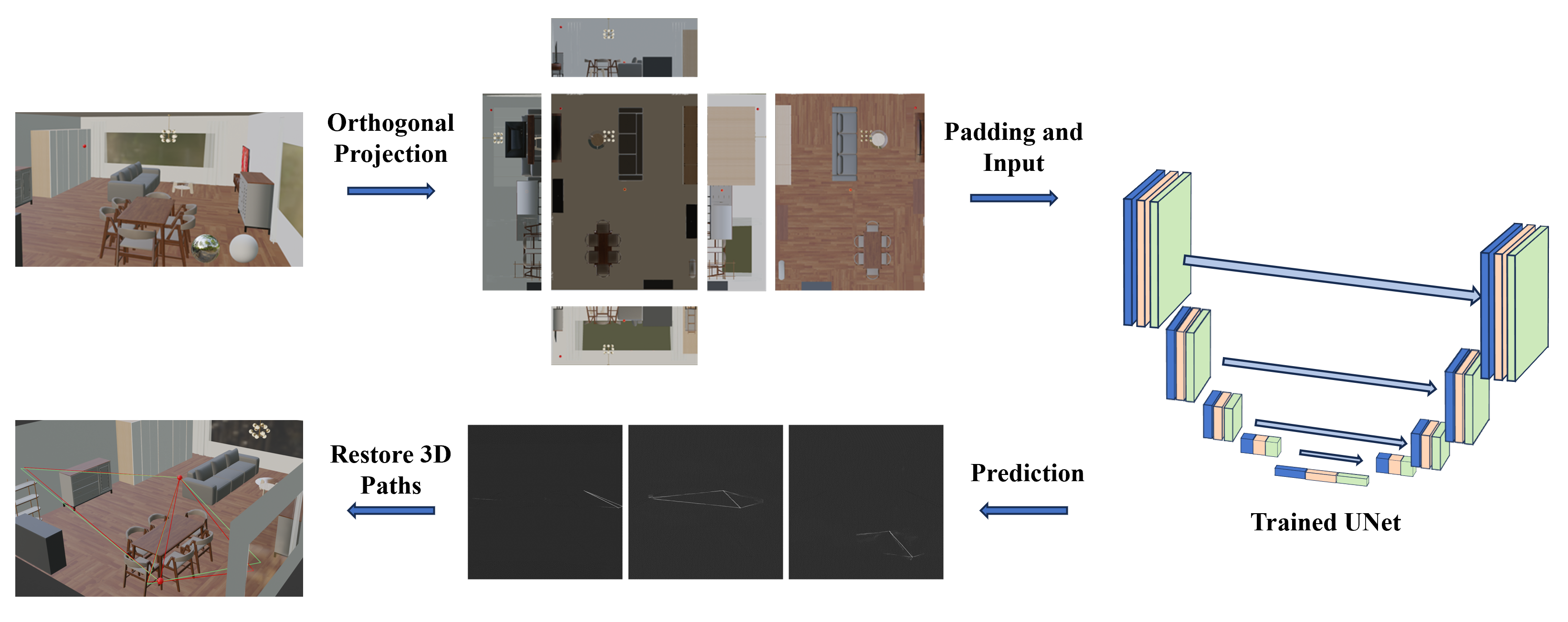}
    \caption{Neural radio tracing to enable direct spatial-CSI prediction from images: A 3D input is projected to six views, then padded and inputted into a trained U-Net network, which predicts the 3D ray paths from three views. This network is able to implicitly learn radio propagation and interactions with objects.}
    \label{fig:radio_tracing}
\end{figure*}

\subsection{Experiment Setting}
\par In this experiment, we aim to evaluate whether a neural network can effectively learn the correlation between the radio environment (including environmental information, Tx and Rx positions) and the spatial paths.  We utilized a generative network to explore this concept, with the complete workflow illustrated in Fig. \ref{fig:radio_tracing}.

\par For dataset, we employed the synthetic WiSegRT dataset\cite{wiseg}, a wireless communication dataset based on realistic, finely modeled indoor environments and ray tracing. This dataset includes ten scenarios, each featuring 15 to 30 Tx locations and correlating to thousands of Rx positions. To simplify the task and facilitate efficient learning, we selected 130,000 Tx-Rx pairs from three indoor scenes and filtered out paths with over two interactions or low channel gain. Then, we constructed the input radio scenes, with orthogonal projection images from six directions $(+x, -x, +y, -y, +z, -z)$. These images were padded to $1024 \times 1024$ pixels to contain enough environment information while maintaining the same scale. Thus, the actual input consists of six $1024 \times 1024$ RGB images, totaling 18 channels. The output consists of three orthogonal projection images of the paths, sufficient for reconstructing these paths in 3D space. This results in a target of three $1024 \times 1024$ gray scale images, totaling three channels. Regarding the division of training, validation, and test sets, considering our goal to validate the feasibility of this concept, we divided them only based on different Tx locations without setting up zero-shot scenes in validation and testing sets. In ideal case, the data should come from hundreds of different environments which are divided into those sets based on environment to test the network's generalization capability.

\par In terms of neural network, we used UNet as the architecture \cite{UNet_ronneberger2015u}, which is widely applied in many generative network architectures due to its encoder and decoder structure with skip connections at multiple network depths. This structure allows for feature extraction from the input domain and restoration to the output domain at different spatial scales. In constructing the convolutional blocks, we used ResNet-18 as the backbone for the encoder part. ResNet \cite{resnet_he2016deep} is an extensively used model for visual information feature extraction, remains a baseline for performance comparisons in many works. The decoder was built using basic deconvolutional layers. Given that the output is gray scale images, binary cross-entropy (BCE) loss function is selected.

\par The training of the network was executed on a server equipped with dual A6000 GPUs. However, even with combined GPU memory of 96 GB, the image size of $1024\times1024$ limits the batch size to 32. Given the dataset's scale, with nearly 5000 batches per epoch, we set an initial learning rate of 0.0001 in the Adam optimization algorithm to avoid the gradient decent falling into a local optimum in the high-dimensional parameter space in early epochs. The subsequent training results have validated the effectiveness of this setting.

\begin{figure*}[ht]
    \centering
    \includegraphics[width=1.0\linewidth]{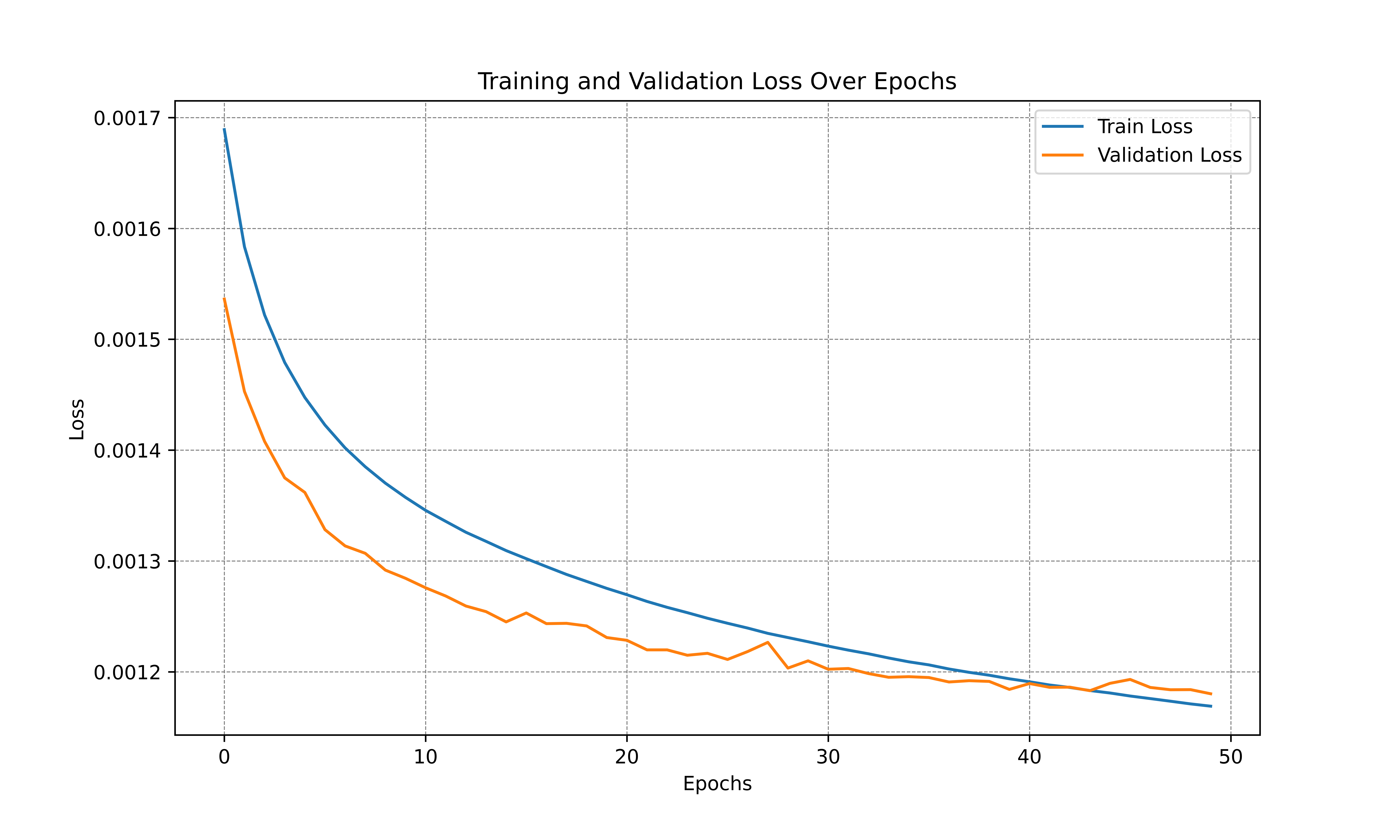}
    \caption{Training and validation loss of the proposed model for spatial-CSI prediction}
    \label{fig:loss}
\end{figure*}

\begin{figure*}[ht]
    \centering
    \includegraphics[width=1.0\linewidth]{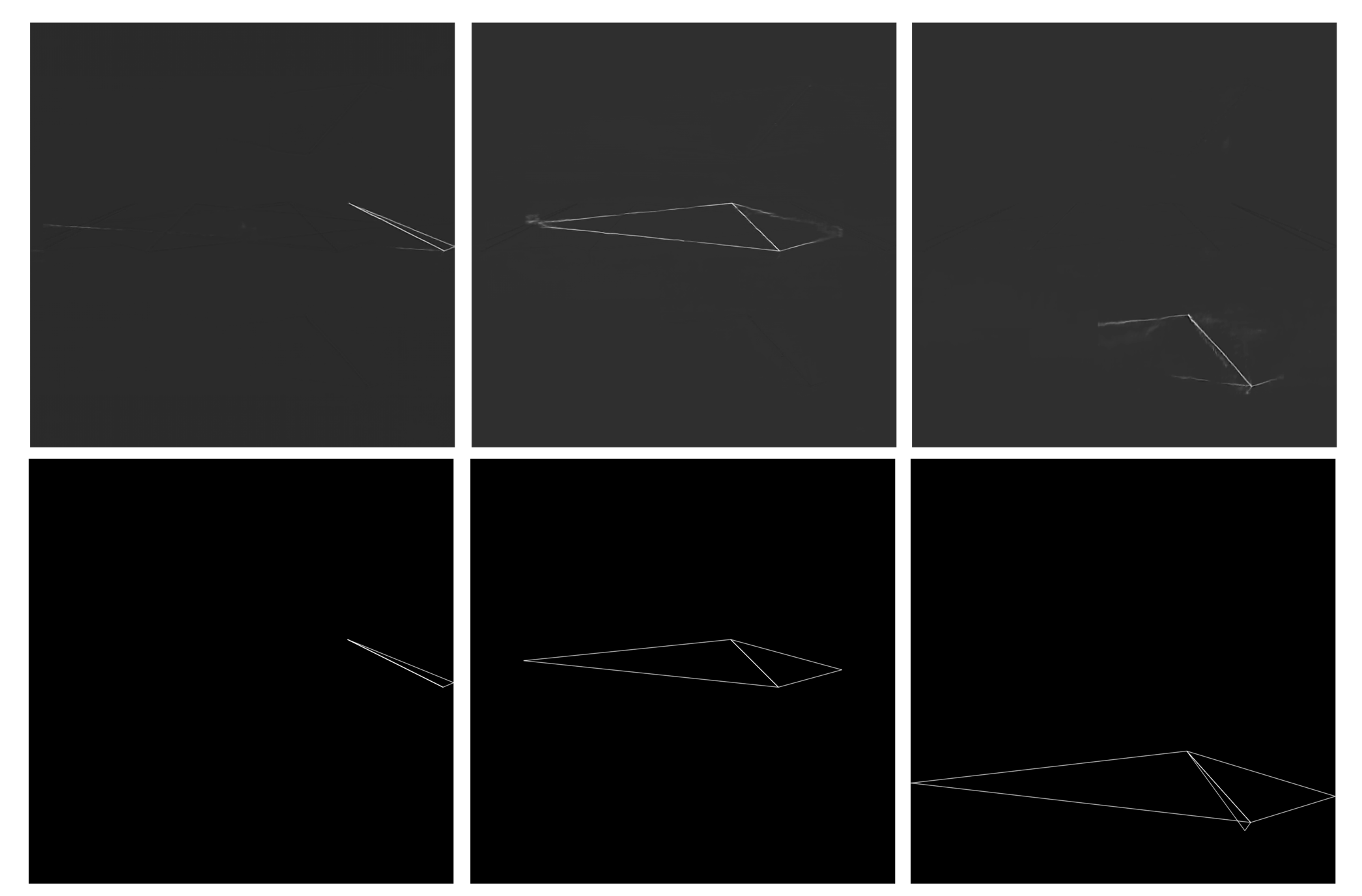}
    \caption{Sample results of the predicted ray paths. Top: predicted results; Bottom: Ground-truth from ray tracing simulator. Background is voided for better illustration. Although blurred and missing some parts, the predicted rays are well-aligned with true spatial paths. Both training and validation data are restricted to two ray interactions.}
    \label{fig:raw_pred}
\end{figure*}

\begin{figure*}[ht]
    \centering
    \includegraphics[width=1.0\linewidth]{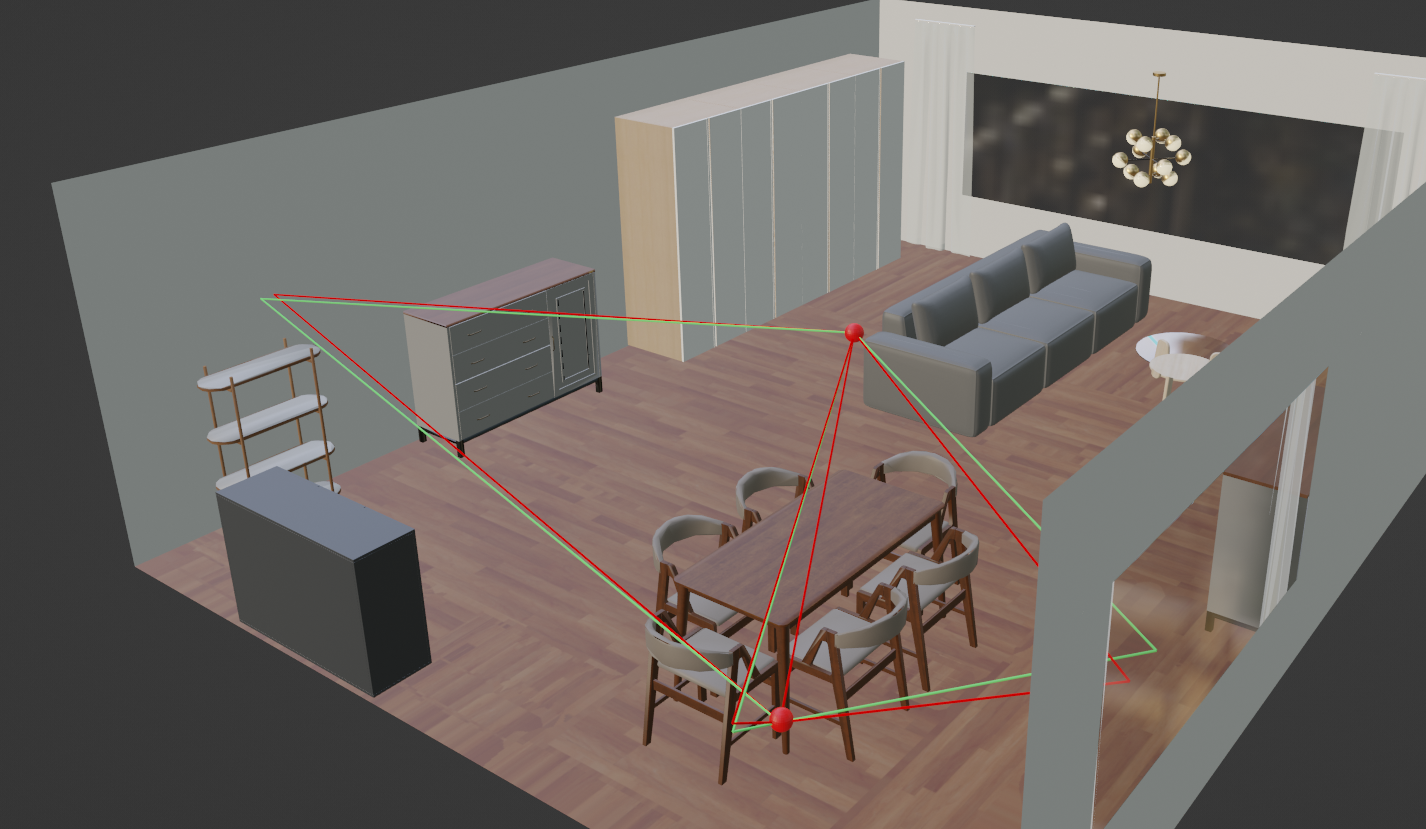}
    \caption{Overlay the predicted paths back to the simulated 3D indoor environment. Red: ground-truth rays; Green: Predicted rays. LoS paths are fully overlapped.}
    \label{fig:recoverd_rays}
\end{figure*}

\subsection{Results}
\par Fig. \ref{fig:loss} shows both the training and validation loss consistently decreased as the training progressed. Around the 50 epochs, the validation loss began to fluctuate, while the training loss continued to decrease, indicating a potential overfitting. Consequently, we chose the model with the lowest validation loss before the 50th epoch. The average test loss was measured at 0.001183, which is close to the validation loss. In terms of the model's performance in predicting paths, as depicted in Fig. \ref{fig:raw_pred}, it is clear that the model has successfully learned the general propagation patterns within the given scenarios. However, there is still a lack of precision in  depicting the paths. Despite this, it was feasible to reconstruct reasonably accurate 3D paths with following simple process. Initially, edge detection algorithms are employed to identify clear paths. Since each image provides positional information in two dimensions, binding these paths across the three orthogonal projections was relatively straightforward. Paths that are clear in all three perspectives are averaged, those blurring in one perspective are compensated by the other two, and those blurring in two perspectives are filled by averaging the nearest endpoints to complete the missing dimension. Given the simple setting on environment characterization, network building and training, achieving such results demonstrates the feasibility of using generative models to understand the environment and identify effective paths. The reconstructed path can be overlaid back to 3D scene, shown in Fig. \ref{fig:recoverd_rays}.

\section{Challenges and Future Opportunities}

\par In our previous discussions, we highlighted how the surge in technological advancements across computation, sensing, and wireless communication hardware, and artificial intelligence fields has presented us with an ambitious vision. This vision involves gathering comprehensive environmental data to construct a real-time digital radio twin, aiming to achieve holographic communication for complete deterministic control over wireless communications. However, numerous significant challenges and opportunities still lie ahead on the path to realizing this vision, which are waiting to be researched and addressed. Here are some foreseeable challenges:

\subsection{Methods for 3D Environmental Reconstruction in Radio Simulation}
Although there are many mature and practical workflows for 3D environmental reconstruction based on multimodal environment data, most of these methods are designed for general tasks. Consequently, some details may lead to a significant decrease in the performance of subsequent radio simulations. For instance, many 3D reconstruction methods output models with uneven surfaces that should be flat. While this might not pose a significant issue for tasks like planning that focus on bounding boxes, it can significantly impact radio simulations. The uneven surfaces reconstructed at angles different from the ground truth can cause a doubled deviation in angle of the reflected wave, thus significantly affecting the accuracy of subsequent propagation. Ensuring the accuracy of the reconstructed plane angles will be crucial.

\subsection{Generalizable Neural Radio Tracing}
In this article, we conducted a simple proof-of-concept validation for employing generative network on neural radio tracing. However, both in terms of accuracy and generalizability, there is a long distance to go before achieving truly practical neural radio tracing. Another challenge is calculating the channel characteristics after identifying the paths. Identifying the factors that most significantly affect the discrepancy between simulation and reality will require considerable effort in field measurement and the modeling of those factors.

\subsection{High-Level Applications}
With the aid of holographic communication and digital radio twins, constructing high-level applications to maximize the benefits is a promising avenue. For example, in the context of CAVs and intelligent transportation systems, the features extracted by the on-board computation units from the data collected by sensors can be fed back to BSs to assist in the construction and updating of scenes. Simultaneously, the path and motion planning of CAVs can help the digital radio twin system to predict channels in advance and make resource reservations for critical services.

\section{Conclusions}

\par In this paper, we investigated the advanced 6G radio technologies, focusing on their exploration of spatial wireless resources. To develop an efficient management scheme on them, we emphasized the importance of spatial-CSI prediction. With considerable overlap to the task of spatial-CSI prediction, this paper proposed the digital radio twin framework. Then we explored the potential of real-time spatial-CSI prediction with generative AI, and proposed a work-flow based on a trained generative network. While our results demonstrate the promise of this approach, it is evident that substantial development is needed before this method can be practically applied in real-world scenarios.

\bibliographystyle{IEEEtran}

\bibliography{lib}
\end{document}